A metabolomic measure of energy metabolism moderates how an inflammatory miRNA relates to rs-fMRI network and motor control in football athletes

Running Title: Tridecenedioate, miR-505, rsfMRI, and balance

Sumra Bari[1,9], Nicole L. Vike[1,9], Khrystyna Stetsiv[1], Linda Papa[2,10], Eric A. Nauman[3,4,5,10], Thomas M. Talavage[3,6,10], Semyon Slobounov[7,10,*], Hans C. Breiter[1,8,9,*]

[1]Department of Psychiatry and Behavioral Sciences, Feinberg School of Medicine, Northwestern University, Chicago, IL, USA
[2] Department of Emergency Medicine, Orlando Regional Medical Center, Orlando, FL, USA
[3]Weldon School of Biomedical Engineering, Purdue University, West Lafayette, IN, USA
[4]School of Mechanical Engineering, Purdue University, West Lafayette, IN, USA
[5]Department of Basic Medical Sciences, Purdue University, West Lafayette, IN, USA
[6]School of Electrical and Computer Engineering, Purdue University, West Lafayette, IN, USA
[7]Department of Kinesiology, Pennsylvania State University, University Park, PA, USA; [8]Laboratory of Neuroimaging and Genetics, Department of Psychiatry, Massachusetts General Hospital and Harvard School of Medicine, Boston, MA, USA

[9,10] indicate co-equal authorship

* Corresponding Authors:
For project design, management and data collection: Semyon Slobounov (sms18@psu.edu)
For hypotheses and conceptual framework, data analysis and paper development: Hans Breiter (h-breiter@northwestern.edu)

**Abstract**

Collision sports athletes experience many head acceleration events (HAEs) per season. The effects of these subconcussive events are largely understudied since HAEs may produce no overt symptoms, and are likely to diffusely manifest across multiple scales of study (e.g., molecular, cellular network, and behavior). This study integrated resting-state fMRI with metabolome, transcriptome and computational virtual reality (VR) behavior measures to assess the effects of exposure to HAEs on players in a collegiate American football team. Permutation-based mediation and moderation analysis was used to investigate relationships between network fingerprint, changes in omic measures and VR metrics over the season. Change in an energy cycle fatty acid, Δtridecenedioate, moderated the relationship between 1) ΔmiR-505 and DMN fingerprint and 2) the relationship between DMN fingerprint and worsening ΔVR Balance measures (all $p_F^{perm}, p_{\beta_3}^{perm} \leq 0.05$). In addition, the similarity in DMN over the season was negatively related to cumulative number of HAEs above 80G, and DMN fingerprint was less similar across the season in athletes relative to age-matched non-athletes. ΔmiR-505 was also positively related to average number of HAEs above 25G per session. It is important to note that tridecenedioate has a double bond making it a candidate for ROS scavenging. These findings between a candidate ROS-related metabolite, inflammatory miRNA, altered brain imaging and diminished behavioral performance suggests that impact athletes may experience chronic neuroinflammation. The rigorous permutation-based mediation/moderation may provide a methodology for investigating complex multi-scale biological data within humans alone and thus assist study of other functional brain problems.

## Introduction

Athletes in impact sports may experience hundreds of so-called "subconcussive" events during a season that do not lead to overt concussive symptoms [1–4], but may lead to subtle alterations in brain structure and function [5–10]. Such events involve a direct blow to the head or an indirect acceleration or whiplash movement due to an impact elsewhere on the body, and are referred to as head acceleration events (HAE) [2]. Neuroimaging studies using functional MRI (fMRI), diffusion weighted imaging, perfusion imaging, and magnetic resonance spectroscopy (MRS) have shown alterations in functional connectivity, working memory, cerebro-vascular reactivity, white matter integrity, regional cerebral blood flow, and neurometabolite concentrations in asymptomatic collision sport athletes due to repeated HAE exposure [8,9,11–19] supporting multiple etiologic hypotheses for injury. Many etiological hypotheses about the mechanism of injury have been proposed, including: *(1) neurovascular decoupling* [20–27] *(2) neuroinflammation* [28–34] and *(3) diffuse axonal injury* (DAI) [35–42]. Although these three models are independently supported by strong evidence, vascular function is a fundamental part of the initiation phase of inflammatory response, and its longer-term resolution phase [43,44], suggesting two of these three models have some overlap.

The neuroinflammation model was the focus of a recent study on micro-RNA (miRNA) levels before and over the course of the football season for miRNAs related to inflammatory responses, such as miR-505, miR-30d, and miR-92a [45]. The panel of miRNAs studied had previously been shown to be elevated in subjects visiting emergency rooms for mild to severe traumatic brain injury (TBI), and were correlated with abnormal clinical readings on CT scans [46]. miRNAs represent a dynamic measure of gene function and are part of the transcriptome [47], targeting up to hundreds of mRNAs, and thus being involved in a wide variety of cellular processes, including many of those that occur after the initial physical impact in head injuries, such as the initiation of inflammation, or its longer-term resolution [48,49]. Inflammation is energy intensive and also leads to alterations in metabolomic profiles (i.e., individualized biochemistry) in athletes, as suggested by several recent MRS studies of mild traumatic brain injury (mTBI) [11,18,50–59]. In parallel, the DAI model suggests there may be an array of shock-wave abnormalities to axons and their structure, and given axon structure depends on a broad array of fatty acids (FAs), FAs might be an important focus for metabolomic study of HAE. One of the fundamental strengths of studying miRNA associated with neuroinflammation, or metabolomic measures of molecules important for brain structural integrity or energetics, is that they can be readily sampled in peripheral blood [45,46]. This strength is coupled, in turn, with the parallel requirement that transcriptomic and metabolomic measures be connected to brain measures and the computational behavior mediated by the brain that is relevant to HAEs and concussion. In recent work, we found such connections between omic measures, brain imaging and computational behavior could be quite powerful, specifically with an observation that miR-30d and miR-92a mediated more than 70% of the relationship between brain imaging and a virtual reality (VR) measure of motor control in football athletes [60].

Core features of the diagnosis of concussion and accumulated HAEs are disturbances of motor control in the form of coordination, balance, or the capacity to navigate, features that were a focus of research by Alexander Luria [61–63]. He noted that head impacts without clinical signs of trauma may be associated with chronic neurocognitive deficits in (a) spatial orientation and accuracy of navigation, (b) processing of visual-spatial information (sensory-motor reactivity), or (c) related coordination functions such as whole-body postural control or balance [61–63]. Recent work with virtual reality (VR) technology has allowed these deficits to be tested using a validated methodology [64,65], producing computational behavior measures of (i) spatial navigation accuracy ("Spatial Memory"), (ii) sensory-motor reactivity and efficiency of visual-spatial processing ("Reaction Time"), and (iii) postural stability during equilibrium changes ("Balance"), along with (iv) an integrative metric of all three measures together ("Comprehensive" score). Although a number of computer-based neuropsychological tests exist to assess aspects of Luria's triad [i.e., (a) – (c) above], VR can detect residual abnormalities in the absence of self-reported symptoms by HAE recipients [66].

Connecting alterations in measures of motor control to in vivo brain alterations requires use of brain imaging such as resting-state fMRI (rs-fMRI), which measures the spontaneous neural activity in the brain and determines the default functional connectivity between brain regions. rs-fMRI has gained wide-spread attention and is used to investigate brain functional connectivity in the normal healthy brain [67–71] as well as in many clinical populations including mTBI [9,12,13,72–74]. A functional connectome (FC) is a symmetric square matrix that estimates the level of functional coupling of blood oxygenation level dependent (BOLD) signal between pairs of brain regions. Robust individual differences in FCs have been termed a "fingerprint", and are demonstrated by the self-identification of subjects by correlating repeat visits of the same subject [75–81]. Use of an association between individual differences in FCs (i.e., fingerprints) and motor control behavior would provide a fundamental way to assure that alterations in transcriptomics and metabolomics were related to HAEs, and not just inflammation based on orthopedic injury.

Given this background, this study evaluated the hypothesis that omic measures from the transcriptome (miRNA) and metabolome (individualized biochemistry) would mediate or moderate the relationship of rs-fMRI measures to motor control behavior. We hypothesized these effects would be observed through three-way associations with computational behavior measures across a season (i.e., pre- vs. post-season) of a football team. We predicted omic measures would show mediation/moderation relationships with imaging and behavioral measures, in line with our prior experience with imaging omics [60]. For the miRNA measures, we focused on a panel of miRNAs that had previously been shown to be abnormal relative to emergency room controls in the pre-season, and across season for football players, and that were known to be involved in the control of inflammation [45]. For metabolomics, we focused only on compounds significantly altered over the season (with FDR correction and/or random forest assessment) that were (i) fatty acids and/or compounds involved with energy metabolism, (ii) compounds involved with stress/inflammatory responses, or (iii) exogenous compounds related to consumption. For rs-fMRI, a network fingerprint approach was used to compare rs-fMRI measures across the season. The broad focus of this imaging-omics study was to determine if identified metabolomic compounds and inflammatory miRNAs mediated or moderated the relationship of rs-fMRI measures to motor control behaviors. Such a finding would go beyond the idea of an intermediate phenotype in "imaging genetics" [82], wherein brain imaging acts as a common, overlapping node between two associations, and point to the quantitative integration of measures across multiple scales, using omics as opposed to genetic measures (that cannot be easily manipulated for clinical purposes). Omic measures such as metabolomics can be clinically manipulated, and their linkage to imaging and computational behavior suggests the possibility for mechanistic insight. To date, we do not believe that imaging omics using metabolomic measures has been described in the literature. To do this, given the large number of associations needing testing for this study, we integrated permutation-based statistics with mediation/moderation analyses to directly address the issue of multiple comparisons.

## Methods

### Participants and data collection

Twenty-three (23) male collegiate varsity student football athletes (mean± standard deviation = 21±2 years) participated in this study. All participants provided informed consent, as approved by the Penn State University Institutional Review Board and in alignment with the Declaration of Helsinki. No participants were excluded due to history of concussion. Self-reports of prior history of concussion (and associated counts) were given by 9 of 23 football athletes (1 prior concussion: n = 7; 2 prior concussions: n = 2). Blood samples were taken prior to any contact practices (*Pre*) and within one week of the last game (*Post*). None of the subjects had a diagnosed concussion in the 9 months preceding preseason data collection. Blood samples were prepared and sent out for miRNA quantification and metabolomic analysis. Concurrent with blood collection, players also underwent virtual reality (VR) testing and MR imaging sessions. HAE measures were collected over the season (i.e., between *Pre* and *Post* blood, imaging, and VR sampling) as described below.

### Head Acceleration Events (HAEs) monitoring

Head acceleration events (HAEs) were monitored at all contact practice sessions (max = 53; no games were monitored) using the BodiTrak sensor system from The Head Health Network. Sensors were mounted in each active player's helmet prior to contact. Sensor outputs included peak translational acceleration (PTA; G-units) and impact location. Two G-unit thresholds i.e. 25G and 80G were selected based on previous reports of impacts related to brain health and injury [83]. For each *i*-th athlete the HAEs were quantified as cumulative number of hits exceeding the threshold *Th* = 25G and 80G ($cHAE_{25G,i}$ and $cHAE_{80G,i}$):

$$cHAE_{Th,i} = \sum_{k=1}^{N} u\left(PTA_{k,i} - Th\right)$$

$$where\ u(x) = \begin{cases} 1\ if\ x > 0 \\ 0\ if\ x\ \leq 0 \end{cases}$$

The average number of hits exceeding 25G and 80G per session for *i*-th athlete ($aHAE_{25G,i}$ and $aHAE_{80G,i}$) is given as

$$aHAE_{Th,i} = \frac{cHAE_{Th,i}}{sessions_i}$$

**Serum Extraction**

Five mL of blood were drawn from each participant at *Pre* and *Post* sessions. Samples were placed in a serum separator tube, allowed to clot at room temperature, and then centrifuged. Serum was extracted from each tube and pipetted into bar-coded aliquot tubes. Serum samples were stored at -70°C until they were transported to 1) a central laboratory for blinded miRNA batch analysis [45] and 2) Metabolon (Morrisville, NC, USA) for blinded metabolite analysis.

**miRNA quantification**

Serum samples collected at *Pre* and *Post* sessions were used to isolate and quantify levels of RNA. 100 $\mu$L of serum was aliquoted and RNA was isolated using a serum/plasma isolation kit (Qiagen

Inc., Venlo, Netherlands) as per the manufacturer's protocol. RNA was eluted in 20 $\mu$L of DNAse/RNAse-free water and stored at -80°C until further use.

Droplet digital PCR (ddPCR; Bio-Rad Inc., Hercules, CA, USA) was used to quantify absolute levels of nine miRNA (miR-20a, miR-505, miR-3623p, miR-30d, miR-92a, miR-486, miR-195, miR-93p, miR-151-5p) [45]. Prior to ddPCR analysis, RNA was checked for quality using a bioanalyzer assay with a small RNA assay. After quality confirmation, 10 ng of RNA was reverse transcribed using specific miRNA TaqMan assays as per the manufacturer's protocol (Thermo Fisher Scientific Inc., Waltham, MA, USA). Protocol details can be found in [45]. The final PCR product was analyzed using a droplet reader (Bio-Rad Inc., Hercules, CA, USA). Total positive and negative droplets were quantified, and from this, the concentration of miRNA/$\mu$L of the PCR reaction was reported. All reactions were performed in duplicate.

**Metabolomic Analysis**

The remaining serum was sent to Metabolon (Morrisville, NC, USA) for metabolomic quantification. Upon arrival, samples were assigned a unique identifier via an automated laboratory system and stored at -80°C. Samples were prepared for subsequent analyses using an automated MicroLab STAR® system (Hamilton Company, Reno, NV, USA). Proteins were precipitated out of each sample using methanol and a shaker (Glen Mills GenoGrinder 2000), and then centrifuged. The resulting extract was then divided into five fractions for various analyses: 1) two fractions for analysis by two separate reverse phase (RP)/UPLC-MS/MS methods with positive ion mode electrospray ionization (ESI), 2) one for analysis by RP/UPLC-MS/MS with negative ion mode ESI, 3) one for analysis by HILIC/UPLC-MS/MS with negative ion mode ESI, and 4) one reserved for backup. To remove organic solvent, samples were briefly placed on a TurboVap® (Zymark); samples were then stored under nitrogen overnight prior to analyses.

Serum metabolites were quantified using Ultrahigh Performance Liquid Chromatography-Tandem Mass Spectroscopy (UPLC-MS/MS). All methods utilized a Waters ACQUITY UPLC and a Thermo Scientific Q-Exactive high resolution/accurate mass spectrometer interfaced with a heated electrospray ionization (HESI-II) source and Orbitrap mass analyzer operated at 35,000 mass resolution. The sample extract was dried and reconstituted in solvents compatible to each of the listed analyses. Each reconstitution solvent contained a series of standards at fixed concentrations to ensure injection and chromatographic consistency. One aliquot was analyzed using acidic positive ion conditions, chromatographically optimized for more hydrophilic compounds. In this method, the extract was gradient eluted from a C18 column (Waters UPLC BEH C18-2.1x100 mm, 1.7 µm) using water and methanol, containing 0.05% perfluoropentanoic acid (PFPA) and 0.1% formic acid (FA). Another aliquot was also analyzed using acidic positive ion conditions, however it was chromatographically optimized for more hydrophobic compounds. In this method, the extract was gradient eluted from the same afore mentioned C18 column using methanol, acetonitrile, water, 0.05% PFPA and 0.01% FA and was operated at an overall higher organic content. Another aliquot was analyzed using basic negative ion optimized conditions using a separate dedicated C18 column. The basic extracts were gradient eluted from the column using methanol and water, however with 6.5mM Ammonium Bicarbonate at pH 8. The fourth aliquot was analyzed via negative ionization following elution from a HILIC column (Waters UPLC BEH Amide 2.1x150 mm, 1.7 µm) using a gradient consisting of water and acetonitrile with 10mM Ammonium Formate, pH 10.8. The MS analysis alternated between MS and data-dependent $MS_n$ scans using dynamic exclusion. The scan range varied slighted between methods but covered 70-1000 m/z.

Peak analysis was conducted using a bioinformatics system which consisted of four major components: 1) the Laboratory Information Management System (LIMS - a system used to automate sample accession and preparation, instrumental analysis and reporting, and data analysis), 2) the data extraction and peak-identification software, 3) data processing tools for quality control and compound identification, and 4) a collection of information interpretation and visualization tools.

Raw data were extracted, peak-identified, and QC processed using Metabolon's hardware and software. Compounds were identified by comparison to library entries of purified standards. Biochemical identifications were based on three criteria: 1) retention index (RI) within a narrow RI window of the proposed identification, 2) accurate mass match to the library (+/- 10 ppm), and 3) the MS/MS forward and reverse scores between the experimental data and authentic standards. The MS/MS scores were based on a comparison of the ions present in the experimental spectrum to the ions present in the library spectrum. While there may have be similarities between these molecules based on one of these factors, the use of all three data points was utilized to distinguish and differentiate more than 3,300 registered biochemicals.

Peaks were quantified using area-under-the-curve. A data normalization step was performed to correct variation resulting from instrument inter-day tuning differences (i.e. variation between pre and postseason analyses). Specifically, each compound was corrected in run-day blocks by registering the medians to equal one (1.00) and normalizing each data point proportionately. Data were then log-transformed. Of the 3,300+ potential biochemicals, 968 metabolites were analyzed at *Pre* and *Post* sessions. Of these, 161 showed significant, FDR-corrected, increases or decreases between the *Pre* and *Post* sessions ($q$-value < 0.05). Of those 161 metabolites, 40 were selected based on the following criteria: 1) they appeared in the random forest plot (20/40) (see Vike et al., 2020 [84]) and 2) they were hypothesized to change following repetitive HAEs. Based on these criteria, six macromolecule categories were defined: lipids (19/40), energy-related metabolites (5/40), xenobiotics (10/40), amino acids (3/40), carbohydrates (2/40), and nucleotides (1/40) [84]. These 40 metabolites were used in all subsequent analyses.

**Virtual Reality (VR) testing**

Athletes completed a previously validated Virtual Reality (VR) neurocognitive testing with a 3D TV system (HeadRehab.com) and a head mounted accelerometer [8,60,65,66] for *Pre* and *Post* sessions. The test included three modules: Spatial Memory, sensory-motor reactivity or whole body Reaction Time and Balance accuracy. These tasks were based on findings from [61–63] for spatial navigation problems following head injuries in veterans. For the Spatial Memory module participants were shown a randomized virtual pathway including multiple turns to a door along with the return trip. Athletes were instructed to repeat the pathway from their memory using a joystick. The Spatial Memory score was based on correct responses versus the errors.

For the Reaction Time module the participants stood feet shoulder width apart with hands on their hips. They were instructed to move their body in the same direction as the virtual room's movements, and the accelerometer measured response time latency.

For the Balance module, athletes were instructed to hold a modified tandem Romberg position for all trials. The first trial was baseline measure where the virtual room was still. Athletes were scored for the subsequent six trials where the virtual room moved in various directions, and the deviance of individual's alignment with the virtual room was quantified via an accelerometer.

To facilitate interpretation, all scores were scaled such that higher score represents better performance.

In addition to the individual scores from each module, an overall Comprehensive score was calculated by combining the three module scores into a ten-point scale (0 worst and 10 best) [66].

**resting-state fMRI**

Twenty (20) athletes participated in two MRI sessions consisting of a 10-min eyes-closed rs-fMRI scan with echo-planar imaging and the following parameters: time of echo (TE) = 35.8 ms, time of repetition (TR) = 2000 ms, flip angle = 90°, 72 contiguous 2-mm axial slices in an interleaved order, voxel resolution = 2 mm × 2 mm × 2 mm, matrix size = 104 × 104 and 300 total volumes. One high-resolution $T_1$ scan using 3D magnetization prepared rapid acquisition gradient recalled echo (3D MPRAGE) sequence was acquired for registration and tissue segmentation purposes with the following parameters: TE = 1.77 ms, time of inversion (TI) = 850ms, TR = 1700ms, flip angle = 9°, matrix size = 320×260×176, voxel size=1mm×1mm×1mm, receiver bandwidth = 300 Hz/pixel, and parallel acceleration factor = 2.

rs-fMRI data were processed using functions from AFNI [85] and FSL [86,87] using in-house MATLAB code explained in detail in [81]. rs-fMRI BOLD timeseries were processed in the subject's native space and the first four volumes were discarded to remove spin history effects. Structural $T_1$ images were denoised and segmented into gray matter (GM), white matter (WM) and cerebrospinal fluid (CSF) tissue masks. The 4D BOLD timeseries was then passed through outlier detection, despiking, slice timing correction, volume registration, aligned to the $T_1$ structural scan, voxel-wise spatial smoothing within tissue masks, scaled to a maximum (absolute value) of 200, and the data were censored to remove outlier timepoints (with the censoring criteria as in [81]). The timeseries were then detrended using no global signal regression with the following common regressors: (1) very low frequency fluctuations as derived from a bandpass [0.002–0.01 Hz] filter, the six motion parameters and their derivatives, and the voxel-wise local neighborhood (40mm) mean WM timeseries.

For connectivity analysis on a regional basis, the grey matter brain atlas from [88] was warped to each subject's native space by linear and non-linear registration. This brain parcellation consists of 278 regions of interest (ROIs). Note that data from the cerebellum (comprising a total of 30 ROIs) were discarded, because the acquired data did not completely cover this structure for all subjects. This resulted in a final GM partition of 248 ROIs. A functional connectivity matrix (namely the functional connectome; FC) was computed for each rs-fMRI scan through correlation of the mean time series from each of the 248 ROIs. The resulting square, symmetric FC matrices were not thresholded or binarized. Each FC matrix was ordered into seven cortical sub-networks, as proposed by Yeo et al. [89], and an additional eighth sub-network comprising sub-cortical regions was added [90]. These networks were: Visual (VIS), Somato-Motor (SM), Dorsal Attention (DA), Ventral Attention (VA), Limbic System (L), Fronto-Parietal (FP), Default Mode Network (DMN) and subcortical regions (SUBC). The fingerprint for FC and these eight networks was calculated by correlating the submatrix corresponding to each of the networks from *Pre* and *Post* sessions [75,76,81]. The network fingerprint represents the similarity between repeat visits of the same participant.

## Statistical Analysis

To assess changes across the season, "Δ" values were calculated for all the aforementioned measures by subtracting *Pre*-session measures from *Post*. For rs-fMRI networks, the fingerprint—representing the similarity between repeat visits of the same participant—was used instead of Δ values. After quality checks and removing missing data seventeen (17) subjects had both *Pre* and *Post* data for all five measurements. All statistical analyses used the software package R [91]. Analyses involved identification of two-way associations, discovery of their overlap as three-way associations, and mediation/moderation testing within a permutation-based framework.

### Two-way associations

Two-way associations were checked between nine network fingerprints (FC, Vis, SM, DA, VA, L, FP, DMN and SUBC), four ΔVR scores (Balance, Reaction Time, Spatial Memory and Comprehensive), nine ΔmiRNA (miR-20a, miR-505, miR-3623p, miR-30d, miR-92a, miR-486, miR-92a, miR-93p, and miR-151-5p), 40 Δmetabolites (complete list as mentioned in [84] and five HAE metrics (sessions, $cHAE_{25G}$, $aHAE_{25G}$, $cHAE_{80G}$, and $aHAE_{80G}$). Linear regression was run between two variables and outliers were removed based on Cook's distance [92], a robust approach to remove outliers. After outlier removal, linear regressions were re-run and all two-way associations with $p \leq 0.05$ were passed forward for further analysis with three-way associations. For all two-way associations meeting the $p \leq 0.05$ threshold, we report the number of outliers removed, $p$-value, β-coefficient and adjusted $R^2$ ($R^2_{adj}$).

### Three-way associations

All two-way associations with $p \leq 0.05$ were used to build three-way associations.

The measurement matrices $X_t, Y_t, Z_t$ defined below schematize the variables used, such as the network fingerprint, VR scores, miRNAs and metabolites.

$$X_t = \begin{bmatrix} x_{t,1,1} & & x_{t,N,1} \\ \vdots & \dots & \vdots \\ x_{t,1,S} & & x_{t,N,S} \end{bmatrix}, Y_t = \begin{bmatrix} y_{t,1,1} & & y_{t,M,1} \\ \vdots & \dots & \vdots \\ y_{t,1,S} & & y_{t,M,S} \end{bmatrix}, Z_t = \begin{bmatrix} z_{t,1,1} & & z_{t,P,1} \\ \vdots & \dots & \vdots \\ z_{t,1,S} & & z_{t,P,S} \end{bmatrix}$$

where N is the total number of variables in matrix $X_t$, M and P are the total number of variables for matrices $Y_t$ and $Z_t$ respectively. S is the number of participants and the matrices were measured at two time points $t = 1$ represents *Pre* and $t = 2$ represents *Post*-season measurements.

Across season measures for VR scores, miRNAs and metabolites were calculated as

$$\Delta X = X_2 - X_1$$

$$\Delta Y = Y_2 - Y_1$$

$$\Delta Z = Z_2 - Z_1$$

Note that, for rs-fMRI instead of difference measures network fingerprint was computed by correlating the data at two time-points.

In what follows, $\Delta \boldsymbol{y_j} \sim \Delta \boldsymbol{x_i}$ will be used to denote the significant two-way association between any two variables (following the procedure described above).

To quantify three-way associations, the following steps (**A** – **C**, below) were performed

First, two-way associations were performed between all variables $\Delta \boldsymbol{y_j}$ and $\Delta \boldsymbol{x_i}$ from matrices $\Delta Y$ and $\Delta X$:

**Step A**: $\Delta \boldsymbol{y_j} \sim \Delta \boldsymbol{x_i} \begin{cases} \forall\ i \in \{1, \cdots, N\} \\ \forall\ j \in \{1, \cdots M\} \end{cases}$

Second, two-way associations were performed between all variables $\Delta \boldsymbol{z_k}$ and $\Delta \boldsymbol{y_j}$ from matrices $\Delta Z$ and $\Delta Y$:

**Step B**: $\Delta \boldsymbol{z_k} \sim \Delta \boldsymbol{y_j} \begin{cases} \forall\ j \in \{1, \cdots M\} \\ \forall\ k \in \{1, \cdots P\} \end{cases}$

Third, two-way associations were performed between all variables $\Delta \boldsymbol{z_k}$ and $\Delta \boldsymbol{x_i}$ from matrices $\Delta Z$ and $\Delta X$:

**Step C**: $\Delta \boldsymbol{z_k} \sim \Delta \boldsymbol{x_i} \begin{cases} \forall\ i \in \{1, \cdots, N\} \\ \forall\ k \in \{1, \cdots P\} \end{cases}$

Finally, three-way associations between any three variables were formed if all three above steps resulted in significant two-way associations for the common variables, leading to the relation:

$$\Delta \boldsymbol{x_i} \sim \Delta \boldsymbol{y_j} \sim \Delta \boldsymbol{z_k} \sim \Delta \boldsymbol{x_i}$$

**Mediation Analysis**

All three-way associations were tested for mediations with ΔVR scores as the dependent variable (DV). This prior hypothesis was observed in previous studies [60]. The independent variable (IV) and mediator (M) were chosen from network fingerprints, ΔmiRNAs and Δmetabolites. Please see the Supplemental material for a detailed description of the mediation procedure.

### Moderation Analysis

All three-way associations were also tested for moderation. The moderation model proposes that the strength and direction of the relationship between independent variable (IV) and dependent variable (DV) is controlled by the moderator variable (M). The IV, DV and M were chosen from network fingerprints, ΔmiRNAs, Δmetabolites and ΔVR scores. The moderation is characterized by the interaction term between IV and M in the linear regression equation as given below:

$$DV = \beta_0 + \beta_1 IV + \beta_2 M + \beta_3 (IV * M) + \epsilon$$

Moderation is significant if $p_{\beta_3} \leq 0.05$ and $p_F \leq 0.05$, where $p_{\beta_3} \leq 0.05$ indicates that $\beta_3$ is significantly different than zero using a t-test and $p_F$ is the $p$-value associated with the overall F-test for the regression equation suggesting that the overall linear relationship is significant.

### Permutation Testing

To control for the occurrence of false positives due to multiple hypotheses testing, permutation-based moderation analysis was conducted. Permutation tests re-sample observations from the original data multiple times to build empirical estimates of the null distribution for the test statistic being studied [93,94]. Permutation-based tests are especially well-suited for studies with small sample sizes as they estimate the statistical significance directly from the data being analyzed rather than making assumptions about the underlying distribution. First, the test statistic is obtained from the original data set, then the data is randomly permuted multiple (Q) times and the test statistic is computed on each permutated data set. The statistical significance is computed by counting (K) the number of times the statistic value obtained in the original data set was more extreme than the statistic value obtained from the permuted data sets, and dividing that value by the number of random permutations (K/Q) [93].

For this study, permutation-based moderation analysis was performed for all three-way associations following the steps listed below:

1. Moderation analysis was performed by assigning the original data variables $\Delta\boldsymbol{x_i}, \Delta\boldsymbol{y_j}, \Delta\boldsymbol{z_k}$ as IV, DV and M to obtain reference test-statistics: $t_0$ and $F_0$. Only variables that formed three-way associations were considered.
2. Data permutation: values were randomly selected from $\boldsymbol{x_{1,i}}$ and $\boldsymbol{x_{2,i}}$ to assign to $\boldsymbol{x'_{1,i}}$ and $\boldsymbol{x'_{2,i}}$.
3. Across season measures were computed from the permuted dataset $\Delta\boldsymbol{x'_i} = \boldsymbol{x'_{2,i}} - \boldsymbol{x'_{1,i}}$. Similarly, $\Delta\boldsymbol{y'_j}$ and $\Delta\boldsymbol{z'_k}$ were computed.
4. Moderation analysis was performed on the permuted dataset $\Delta\boldsymbol{x'_i}$, $\Delta\boldsymbol{y'_j}$, $\Delta\boldsymbol{z'_k}$ by assigning as IV, DV and M and the test statistics: $t'_q$and $F'_q$ were obtained.
5. The counter variable $K_1$ was incremented by one if absolute value of $t_0$ was greater than absolute value of $t'_q$.
6. $K_2$ was incremented by one if absolute value of $F_0$ was greater than absolute value of $F'_q$.
7. Steps 2-6 were repeated: $q = 1,2, \cdots, Q$ times. Here, $Q = 100{,}000$.

8. Permutation-based $p$-value $p_{\beta_3}^{perm}$ was calculated as the proportion of the $t'_q$ values that are as extreme or more extreme than $t_0$ i.e. $K_1/Q$ .
9. Permutation-based $p$-value $p_F^{perm}$ was computed from $F_0$ and $F'_q$ i.e. $K_2/Q$.
10. Moderation analysis was considered significant if $p_{\beta_3}^{perm} \leq 0.05$ and $p_F^{perm} \leq 0.05$.

## Results

This analysis integrated four types of measure at two time-points *Pre* and *Post* from a competition season of American collegiate football using a permutation-based mediation/moderation framework. The four measures integrated in mediation/moderation were: network fingerprint (reflecting changes in brain connectivity), change in metabolomics, changes in miRNA and changes in VR-based motor control. For such a mediation/moderation framework, two-way associations were identified using Cook's distance assessments for outlier removal, and then three-way associations were identified through the overlap of two-way associations. Mediation testing was done in a directed manner, with motor control behavior always being the dependent variable following prior methods (e.g., as seen by Chen et al., 2019 [60]). Moderation testing looked at all possibilities for the dependent variable. All mediation/moderation testing used a permutation-based framework to provide protection against false positives due to multiple comparisons. Secondary analyses done after permutation-based mediation/moderation included (i) assessment of the brain imaging measures in athletes relative to age-matched controls, and (ii) testing of association between the four measures integrated in mediation/moderation against HAE measures.

## Two-way associations

Linear regression following Cook's outlier removal was used to assess significant ($p \leq 0.05$) two-way associations between functional connectome (FC) and eight network fingerprints (Vis, SM, DA, VA, L, FP, DMN and SUBC), four ΔVR scores (Balance, Reaction Time, Spatial Memory and Comprehensive), nine ΔmiRNA (miR-20a, miR-505, miR-3623p, miR-30d, miR-92a, miR-486, miR-92a, miR-93p, and miR-151-5p), 40 Δmetabolites (complete list as mentioned in [84]. Secondary analyses included assessments of these four categories of measure against the five HAE metrics (sessions, $cHAE_{25G}$, $aHAE_{25G}$, $cHAE_{80G}$, and $aHAE_{80G}$).

Supplemental Table 1 (A-J) report the significant negative or positive pairwise associations between any two variables with number of Cook's outliers removed, $p$-value, β coefficient and adjusted $R^2$ ($R^2_{adj}$) for the regression. The common significant pairwise associations were then used to build three-way associations.

## Three-way associations

Three-way associations were formed from the common significant pairwise associations between any three variables $\Delta \boldsymbol{x_i}, \Delta \boldsymbol{y_j}$ and $\Delta \boldsymbol{z_k}$ where these could be any of the network fingerprints, ΔVR scores, ΔmiRNAs and/or Δmetabolites (see Methods for details).

Table 1 (A-C) lists all three-way associations that resulted from the common pairwise associations as described in Methods, and were used for directed mediation and moderation analysis.

## Mediation and Moderation results

The three-way associations were tested for directed mediation analysis with ΔVR as the dependent variable. The independent variable and mediator were chosen from network fingerprint, ΔmiRNA and/or Δmetabolite. None of the three-way associations resulted in a significant mediation.

The three-way associations identified and listed in Table 1 (A-C) were then tested for moderation analysis across network fingerprint, ΔVR, Δmetabolite and ΔmiRNA. Table 1 (A-C) lists the significant moderation result ($p_F^{perm} \leq 0.05$, $p_{\beta_3}^{perm} \leq 0.05$) in bold and italics. Two moderation results were observed meeting permutation criteria.

**(A)**

| | | | | Path A: Y~X | | Path B: Z~Y | | Path C: Z~X | | | |
|---|---|---|---|---|---|---|---|---|---|---|---|
| **ΔMetabolite (X)** | **Network(Y)** | **ΔVR (Z)** | **Cook's outliers** | $\beta_A$ | $p_A$ | $\beta_B$ | $p_B$ | $\beta_C$ | $p_C$ | $p_{\beta_3}^{perm}$ | $p_F^{perm}$ |
| 2-hydroxyglutarate | SM | Comprehensive | 4/17 | -0.43 | 0.04 | -0.68 | 0.01 | 0.56 | 0.01 | 0.73 | 0.00 |
| 1,7-dimethylurate | SM | Comprehensive | 3/17 | -0.39 | 0.03 | -0.68 | 0.01 | 0.45 | 0.04 | 0.28 | 0.15 |
| 7-hydroxyoctanate | FC | Balance | 1/17 | 0.48 | 0.03 | 0.44 | 0.02 | 0.45 | 0.04 | 0.43 | 0.68 |
| ***tridecenedioate*** | ***DMN*** | ***Balance*** | ***3/17*** | ***-0.26*** | ***0.05*** | ***0.45*** | ***0.02*** | ***-0.44*** | ***0.04*** | ***0.03*** | ***0.05*** |
| 2-hydroxyglutarate | SM | Reaction Time | 2/17 | -0.43 | 0.04 | -0.74 | 0.00 | 0.54 | 0.01 | 0.63 | 0.05 |

**(B)**

| | | | | Path A: Y~X | | Path B: Z~Y | | Path C: Z~X | | | |
|---|---|---|---|---|---|---|---|---|---|---|---|
| **ΔmiRNA (X)** | **ΔMetabolite (Y)** | **Network(Z)** | **Cook's outliers** | $\beta_A$ | $p_A$ | $\beta_B$ | $p_B$ | $\beta_C$ | $p_C$ | $p_{\beta_3}^{perm}$ | $p_F^{perm}$ |
| ***505*** | ***tridecenedioate*** | ***DMN*** | ***4/17*** | ***0.45*** | ***0.05*** | ***-0.26*** | ***0.05*** | ***-0.46*** | ***0.02*** | ***0.02*** | ***0.02*** |

**(C)**

| | | | | Path A: Y~X | | Path B: Z~Y | | Path C: Z~X | | | |
|---|---|---|---|---|---|---|---|---|---|---|---|
| **miRNA (X)** | **Network(Y)** | **VR (Z)** | **Cook's outliers** | $\beta_A$ | $p_A$ | $\beta_B$ | $p_B$ | $\beta_C$ | $p_C$ | $p_{\beta_3}^{perm}$ | $p_F^{perm}$ |
| 505 | DMN | Balance | 2/17 | -0.46 | 0.02 | 0.45 | 0.02 | -0.53 | 0.00 | 0.48 | 0.13 |

**Table 1**. Table (A), (B) and (C) lists all triangulations. $\beta_A, p_A$ are the nominal interaction results between X and Y variables; $\beta_B, p_B$ are the nominal interaction results between Y and Z variables and $\beta_C, p_C$ are the nominal interaction results between X and Z variables. Cook's outliers (*k*/*n*) lists the number of outliers *k*, based on Cook's distance, removed out of the total *n* samples for moderation analysis. Significant moderation results are marked in bold and italics. Table (A) lists the results between ΔMetabolite, Network fingerprint and ΔVR tasks. Table (B) lists the results between ΔmiRNA, ΔMetabolite and Network fingerprint. Table (C) lists the results between ΔmiRNA, Network fingerprint and ΔVR tasks

One significant moderation result is shown in Figure 1(A-C) between ΔVR Balance (dependent variable), and DMN fingerprint, Δtridecenedioate (independent variable, moderator) with $p_F^{perm}$ = 0.03, $p_{\beta_3}^{perm}$= 0.05. The regression slope (β), $p$-value and adjusted R2 ($R_{adj}^2$) depicted on each arm of the triangle corresponds to the pairwise interaction results, after Cook's distance outliers removed, between those two variables with $p \leq 0.05$. Figure 1(B) shows the pairwise interactions between the three variables ΔVR Balance, DMN fingerprint and Δtridecenedioate. There was a negative relationship between DMN fingerprint and Δtridecenedioate, a negative relationship between Δtridecenedioate and ΔVR Balance and a positive relationship between DMN fingerprint and ΔVR Balance. Figure 1(C) plots the relationship between DMN fingerprint and ΔVR Balance with Δtridecenedioate as the moderator. Three lines corresponds to low (mean - standard deviation), medium (mean) and high (mean + standard deviation) Δtridecenedioate values.

A second significant moderation result is shown in Figure 1(D-F) between DMN fingerprint (dependent variable), and ΔmiR-505, Δtridecenedioate (independent variable, moderator) with $p_F^{perm}$ = 0.02, $p_{\beta_3}^{perm}$= 0.02. The regression slope (β), $p$-value and adjusted R2 ($R^2_{\mathrm{adj}}$) depicted on each arm of the triangle corresponds to the pairwise interaction results, after Cook's distance outliers removed, between those two variables with $p \leq 0.05$. Figure 1(E) plots the pairwise interactions between these three variables. There was a positive relationship between ΔmiR-505 and Δtridecenedioate, a negative relationship between DMN fingerprint and Δtridecenedioate and a negative relationship between DMN fingerprint and ΔmiR-505. Figure 1(F) depicts the relationship between Δ miR-505 and DMN fingerprint with Δtridecenedioate as the moderator. Three lines corresponds to low (mean - standard deviation), medium (mean) and high (mean + standard deviation) Δtridecenedioate values.

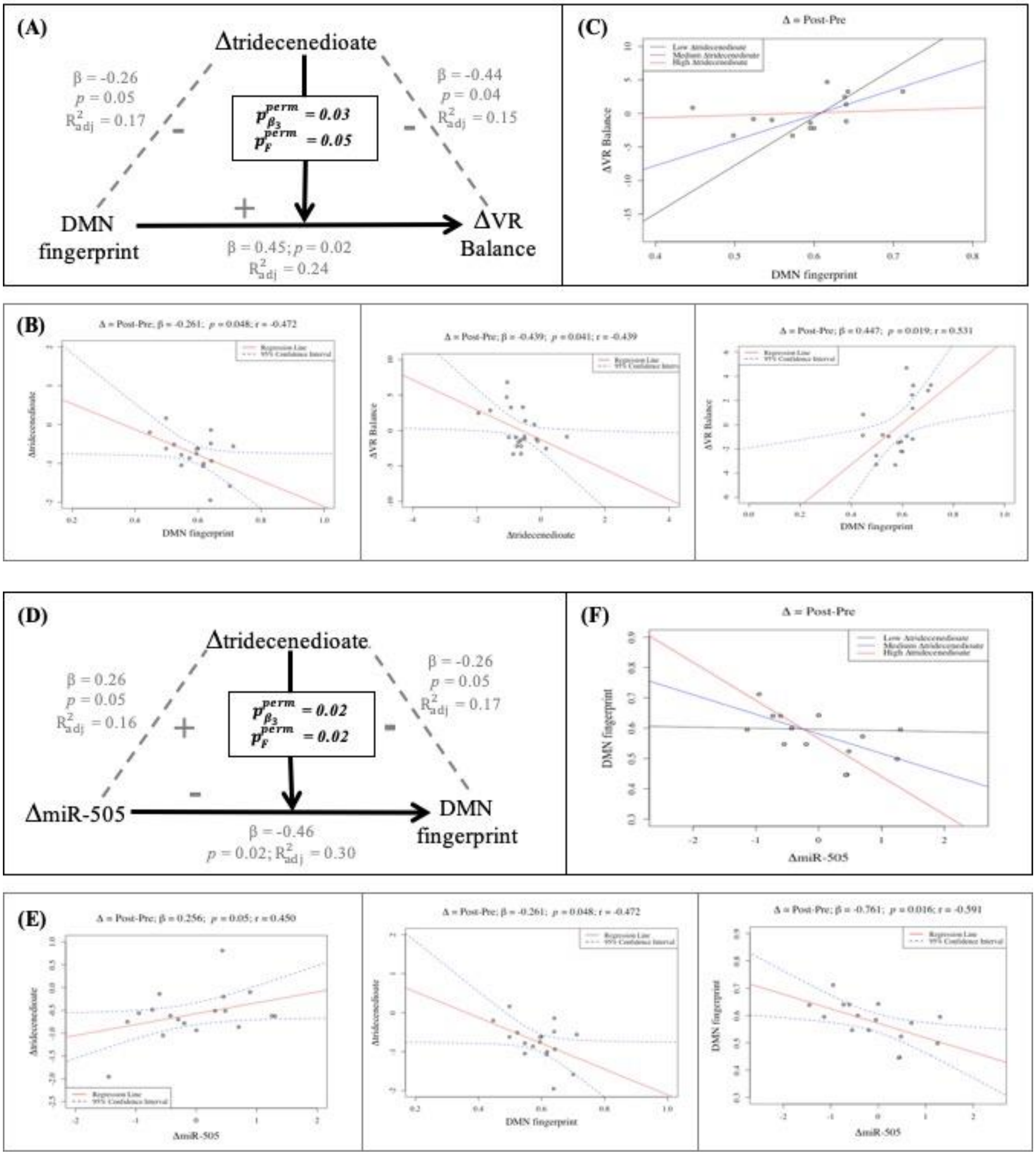

**Figure 1**. **(A-C)** Significant moderation analysis result. **(A)** ΔVR Balance was the dependent variable, DMN fingerprint and Δtridecenedioate were independent variable and moderator ( $p_F^{perm} = 0.05, p_{\beta_3}^{perm} = 0.03$). The regression slope (β), $p$-value and adjusted R2 ($R^2_{adj}$) depicted on each arm of the triangle corresponds to the significant interaction results, after Cook's distance outliers removed, between those two variables with $p$-values reported at a significance level of 0.05. **(B)** Interaction plots corresponding to the moderation analysis. There was negative relationship between DMN fingerprint and Δtridecenedioate; there was negative relationship between Δtridecenedioate and ΔVR Balance; there was positive relationship between DMN fingerprint and ΔVR Balance. **(C)** The plot depicts the relationship between DMN fingerprint and ΔVR Balance with Δtridecenedioate as the moderator.

Three lines corresponds to low (mean - standard deviation), medium (mean) and high (mean + standard deviation) Δtridecenedioate values. **(D-F)** Significant moderation analysis result. **(D)** DMN fingerprint was the dependent variable, and ΔmiR-505 and Δtridecenedioate were independent variable and moderator ($p_F^{perm} = 0.02, p_{\beta_3}^{perm} = 0.02$). The regression slope (β), $p$-value and adjusted $R_2$ ($R^2_{adj}$) depicted arm of the triangle corresponds to the significant interaction results, after Cook's distance outliers removed, between those two variables with $p$-values reported at a significance level of 0.05. **(E)** Interaction plots corresponding to the moderation analysis. There was positive relationship between Δtridecenedioate and ΔmiR-505; there was negative relationship between Δtridecenedioate and DMN fingerprint; there was negative relationship between DMN fingerprint and ΔmiR-505. **(F)** The plot depicts the relationship between DMN fingerprint and ΔmiR-505 with Δtridecenedioate as the moderator. Three lines corresponds to low (mean - standard deviation), medium (mean) and high (mean + standard deviation) Δtridecenedioate values.

Given the same metabolomic measure moderated both results, Figure 2 integrates the two moderation analyses results along with HAE metrics. Figure 2(A) shows that ΔVR Balance was the dependent variable and DMN fingerprint acted as the independent variable for one moderation and dependent variable for the other. Δtridecenedioate acted as the moderator for the two moderation analyses. In addition, there was a positive relationship between ΔmiR-505 and $aHAE_{25G}$ as depicted in Figure 2(B). Figure 2(C) shows the negative relationship between DMN fingerprint and $cHAE_{80G}$.

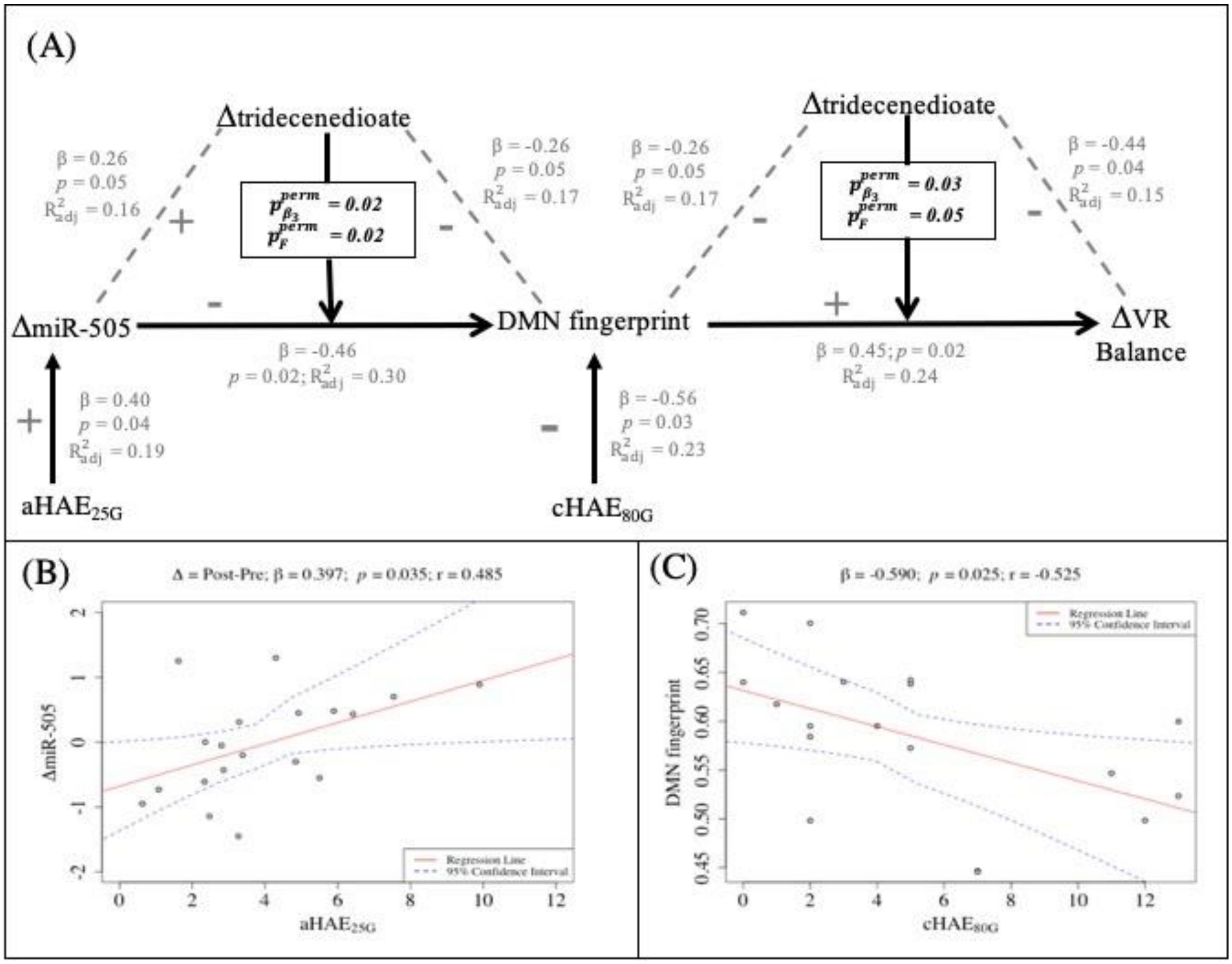

**Figure 2**. This figure combines the moderation analyses results with HAEs. **(A)** ΔVR Balance was the dependent variables and ΔmiR-505 was the independent variable. DMN fingerprint acted as independent variable for one moderation and as dependent variable for the other one. Δtridecenedioate acted as the moderator for the two analyses. The regression slope (β), $p$-value and adjusted R2 ($R^2_{adj}$) depicted arm of the diamond corresponds to the significant interaction results, after Cook's distance outliers removed, between those two variables with $p$-values reported at a significance level of 0.05. This figure depicts the integration of resting state fMRI with metabolic profiling, transcriptomics and computational virtual reality (VR) behavior task along with Head Acceleration Events (HAEs). **(B)** Interaction plots between Network fingerprint, ΔmiRNA with Head Acceleration Events (HAEs). The regression slope (β), $p$-values and correlation coefficient (r) depicted on top of the plots corresponds to the interaction results, after Cook's distance outliers removed, between two variables with $p$-values reported at a significance level of 0.05. There was a positive relationship between ΔmiR-505 and average number of HAEs above 25G per session ($aHAE_{25G}$). **(C)** There was a negative relationship between DMN fingerprint and cumulative number of HAEs above 80G ($cHAE_{80G}$).

## rs-fMRI between Age-Matched Non-Athletes (NAth) and Football Athletes (Ath)

To facilitate interpretation of the permutation-based mediation/moderation analyses, network fingerprint strength was compared between age-match non-athletes (NAth) and football athletes (Ath) in this study.

Figure 3(A) shows the brain rendering of average network fingerprint strength across the cohort for NAth and Ath. The NAth exhibited stronger fingerprint for each of the resting-state fMRI network: Vis, SM, DA, VA, L, FP, DMN and SUBC. Figure 3(B) shows the boxplots for all the network fingerprints for age-match non-athletes and football athletes. Age-matched non-athletes exhibited significantly higher network fingerprint values, using two-sample t-test following Bonferroni correction, for all networks except the Limbic system (L), as compared to the football athletes.

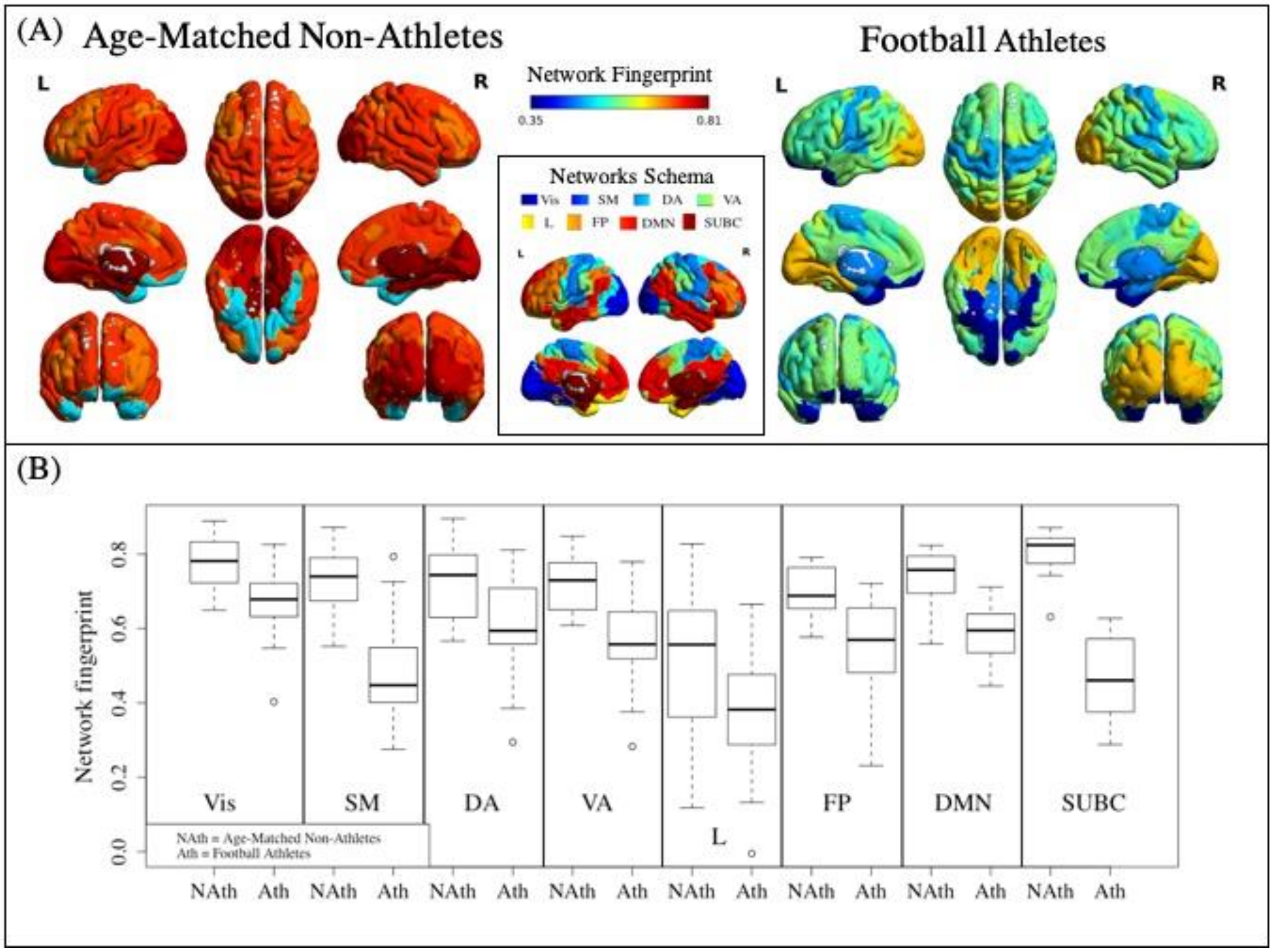

**Figure 3.** Network fingerprint strength analysis for each of Yeo's resting state functional networks between football athletes and age-matched non-athletes. **(A)** Brain rendering of average network fingerprint across football athletes and age-matched non-athletes. The strength per brain region depicted in the figure is the network fingerprint of the cohort to which the brain region belongs. **(B)** Boxplots of network fingerprint for each of Yeo's resting state functional networks between football athletes and age-matched non-athletes. Note that football athletes have significantly lower network fingerprint than the age-matched non-athletes for all networks except Limbic system (L) using two-sample t-test followed with Bonferroni correction.

Yeo's resting functional networks [89]: Visual (VIS), Somato-Motor (SM), Dorsal Attention (DA), Ventral Attention (VA), Limbic system (L), Fronto-Parietal (FP), Default Mode Network (DMN), and subcortical regions (SUBC).

**Discussion**

This study evaluated the hypothesis that omic measures from the transcriptome (miRNA) and metabolome (individualized biochemistry) would mediate or moderate the relationship of rs-fMRI measures to motor control behavior collected *Pre* and *Post* season in collegiate American football athletes. Using a rigorous outlier removal procedure and permutation analyses for multiple comparisons correction, this hypothesis was confirmed for one metabolomic measure showing multiple moderation effects. Specifically, a fatty acid, tridecenedioate, was found to moderate the relationship of the DMN fingerprint from rs-fMRI to motor control (balance) behavior. The DMN network was shown to have less self-similarity over the course of the season in football players relative to age-matched controls, and it showed a negative relationship with HAEs accumulated over the season (i.e., more HAEs meant lower similarity in DMN across the season). Further, HAEs showed a positive association with changes in miR-505 levels, indicating that increasing HAEs were associated with increasing levels of a neuroinflammatory-associated miRNA. Lastly, tridecenedioate moderated the relationship of miR-505 to the DMN network.

Tridecenedioate is a long-chain monounsaturated (i.e., one double bond) dicarboxylic fatty acid. Such compounds are hypothesized to act as reactive oxygen species (ROS) scavengers given the double bond stability and reactivity. Interestingly, similar monounsaturated compounds, as observed [84], were also decreased (e.g., heptenedioate, hexedecenedioate). Long-chain polyunsaturated fatty acids (PUFAs) have anti-inflammatory properties via the reduction of ROS and direct superoxide scavenging [95], while monounsaturated fatty acids (MUFAs) display neuroprotective effects and ameliorate brain injury in animal models [96]. Unsaturated FAs are also shown to efficiently antagonize the toxic action of saturated non-esterified fatty acids [97], and the presence of the double bond corresponds to increased antioxidant and neuroprotective effects [98]. Thus, the association between across-season decreases in tridecenedioate with inflammatory miR-505 (see Supplemental Figure1) implicate its role in ROS scavenging and inflammation resolution.

Overall, the selected miRNAs have been implicated in cancer, systemic inflammation, and central nervous system disorders [99–107]. Specifically, miR-505 is negatively correlated with progression in various cancers, inhibiting tumorigenesis and acting as a tumor suppressor in malignancies such as glioblastoma, endometrial carcinoma, hepatocellular carcinoma and prostate cancer [108–111]. In osteosarcoma tissues, for instance, the reduced expression of miR-505 is significantly associated with poorer clinical prognosis [112]. miR-505 has been reported to inhibit osteosarcoma and hepatoma cell proliferation, migration and invasion by regulating the high-mobility group box 1 (HMGB1), which functions as a damage-associated molecular pattern that propagates infection- or injury-elicited inflammatory responses [112,113]. In addition to its oncogenic associations, miR-505 was found to be significantly decreased in the blood of Ulcerative Colitis patients, thus playing a critical role in chronic inflammatory bowel disease [114]. Furthermore, miR-505 has been implicated in Alzheimer’s disease [115] and identified as a predictive biomarker for Parkinson’s disease (PD) [116]. In PD, miR-505 had a potential functional role via inverse modulation of neural proliferation differentiation and control 1 (NPDC1), a down-regulator of neural proliferation [117]. miR-505 was also shown to be upregulated in both mild to moderate and severe TBI in human subjects [46]. Here, involvement of miR-505 suggests a state of neuroinflammation in asymptomatic football athletes due to repetitive HAE exposure. miR-505 was not only associated with decreased tridecenedioate, but also with decreased self-identifiability in the DMN.

Default Mode Network (DMN) is the resting state fMRI network that appeared in both moderation results. Alterations in this network have been associated with numerous neurological disorders. DMN, an ensemble of cortical regions, typically including parts of the anterior and the posterior cingulate cortices, is a task-negative resting state network that is active while at rest and deactivated during a task. Apart from being active during rest, DMN persists in passive sensory processing tasks with minimal cognitive demands [70,118,119] but switches off during externally cued tasks with high cognitive demands. DMN is essential for basic, and possibly subconscious, processing necessary for calibrating affective and autonomic states of the brain [70]. DMN is also critical for retrieval of episodic memory [70,120–123] and altered connectivity has been observed in patients with Alzheimer's disease [70,124–128]. Altered DMN connectivity, and the subsequent inability to deactivate it for cognitive tasks, has also been observed in patients with schizophrenia [129,130], depression [131–133], epilepsy [134–136], multiple sclerosis [137], Autism Spectrum Disorder [138] and contributes to executive deficits in Parkinson Disease (PD) patients [139] such as PD-associated saccadic hypometria [140] and altered resistance to passive movement/motor performance in PD patients [141,142]. Altered DMN connectivity has also been observed in concussed and TBI patients [12,143–151]. Despite these profound network changes, relating complex neuroimaging such as rs-fMRI to observable behavior remains elusive.

Core features of the diagnosis of concussion and accumulated HAEs over a season, are disturbances of motor control in the form of coordination, balance, or the capacity to navigate as investigated by Alexander Luria [61–63]. Numerous studies have attempted to uncover behavioral changes in contact sport athletes, but many tests are insensitive to subtle changes in behavior [152–156]. Recent advances in virtual reality (VR) has prompted the investigation of these deficits in concussed athletes with the use of a validated methodology [64,65]. In order to observe subtle changes in contact sports athletes with or without concussion, it is critical to study associations between motor control behavior and other measures, such as metabolomics and brain imaging. In this study, changes in VR Balance scores over the season were found to be associated with DMN fingerprint and Δtridecenedioate. Better performance in the VR Balance task was related to higher DMN similarity (i.e., higher fingerprint values) over the season and lower levels of the ROS scavenging compound, tridecenedioate. This implied that improved motor control across the season of play was associated with smaller changes in the brain's executive network and more efficient neuroinflammation resolution. Combining complex computational VR tasks with other indicators of neurological dysfunction (i.e., metabolomics and neuroimaging) proved to be a powerful method to uncover subtle, and unobservable changes in brain function related to repetitive HAEs.

DMN fingerprint and ΔmiR-505 were also found to be correlated with the HAE metrics. First, the DMN fingerprint, which depicts similarity between repeat visits for a specific participant, was found to be significantly lower in football athletes as compared to age-matched non-athletes. This implies that the DMN network, which is critical for executive functions, is altered in football athletes. Specifically, the fingerprint was negatively associated with the cumulative number of high magnitudes hits; namely, the higher the number of $cHAE_{80G}$, the lower the similarity in DMN across the season. Second, ΔmiR-505 was found to be positively associated with $aHAE_{25}$; therefore, elevated levels of an inflammatory miRNA were positively associated with a higher number of average HAEs over the season. One interpretation of these results is that limiting HAE exposure over the season could potentially mitigate the alterations observed in these asymptomatic athletes.

The primary limitation of the study was the modest sample size for which all five measures were available: transcriptome, metabolome, MR imaging, VR behavior, and HAE data. It is critical to note that with the highly rigorous mathematical approach used in this study to analyze the dataset, results are deemed highly reproducible despite this potential limitation. Another potential limitation of the study was that the HAE metrics were only allowed to be collected at practices and not during competitive play, which resulted in a limited picture of the true exposure to "subconcussive" events. To observe the transient changes that football athletes go through throughout a competition season, these metrics would need to be collected over more impact events in future studies. Future studies might also include comparison of the aforementioned metrics collected for football athletes to age-matched, non-contact sports athletes.

## Conclusion

This study evaluated the associations between metabolome, transcriptome, brain imaging, and behavior measures, collected in collegiate football athletes before and after the season, as a function of mechanical accelerations to the head. It was found that Δtridecenedioate moderated the relationship between 1) ΔmiR-505 and DMN fingerprint and 2) the relationship between DMN fingerprint and ΔVR Balance measures. Metabolomics provides a viable method for investigating metabolic fluctuations and biomarkers of brain injury [157,158], but often lack a relationship to neurological function or dysfunction (observed with brain imaging or behavioral testing) and other markers of neuroinflammation (i.e. miRNAs). Here, a metabolomic measure was found to be associated with the transcriptome, brain imaging, and behavior, elucidating it as a potential HAE-associated blood biomarker of neurological dysfunction. The integration of these multi-scale biological measures through a permutation-based approach uncovered a mechanism of putative brain injury due to HAEs as summarized in Figure 4 and described in detail in [84]. The findings of this study are synergistic with the findings of Vike and colleagues in that markers of mitochondrial distress were linked to HAEs, and metabolites related to fatty acid oxidation (e.g. sebacate) and the TCA cycle (e.g. citrate) were observed to mediate the relationship between miRNAs and behavior [84]. Mitochondrial dysfunction is suspected to increase reactive oxygen species production, which is supported by the alterations we observed with $\Delta$tridecenedioate, which was in turn linked to chronic elevations in inflammatory miRNAs and alterations in brain imaging and behavior. The rigorous permutation-based mediation/moderation that integrated omic measures with brain imaging and behavior suggests a viable methodology for investigating complex multi-scale biological data. Altogether, this study suggests 1) a potential mechanism of brain injury due to repetitive HAEs and 2) a permutation-based mediation/ moderation approach to integrate multi-scale data that can be applied broadly. This last point is of importance in the context of current controversies about the relevance of animal models of functional brain illness, such as with mental illness, and suggests one possible route forward for developing mechanistic hypotheses from the study of humans alone [159–165].

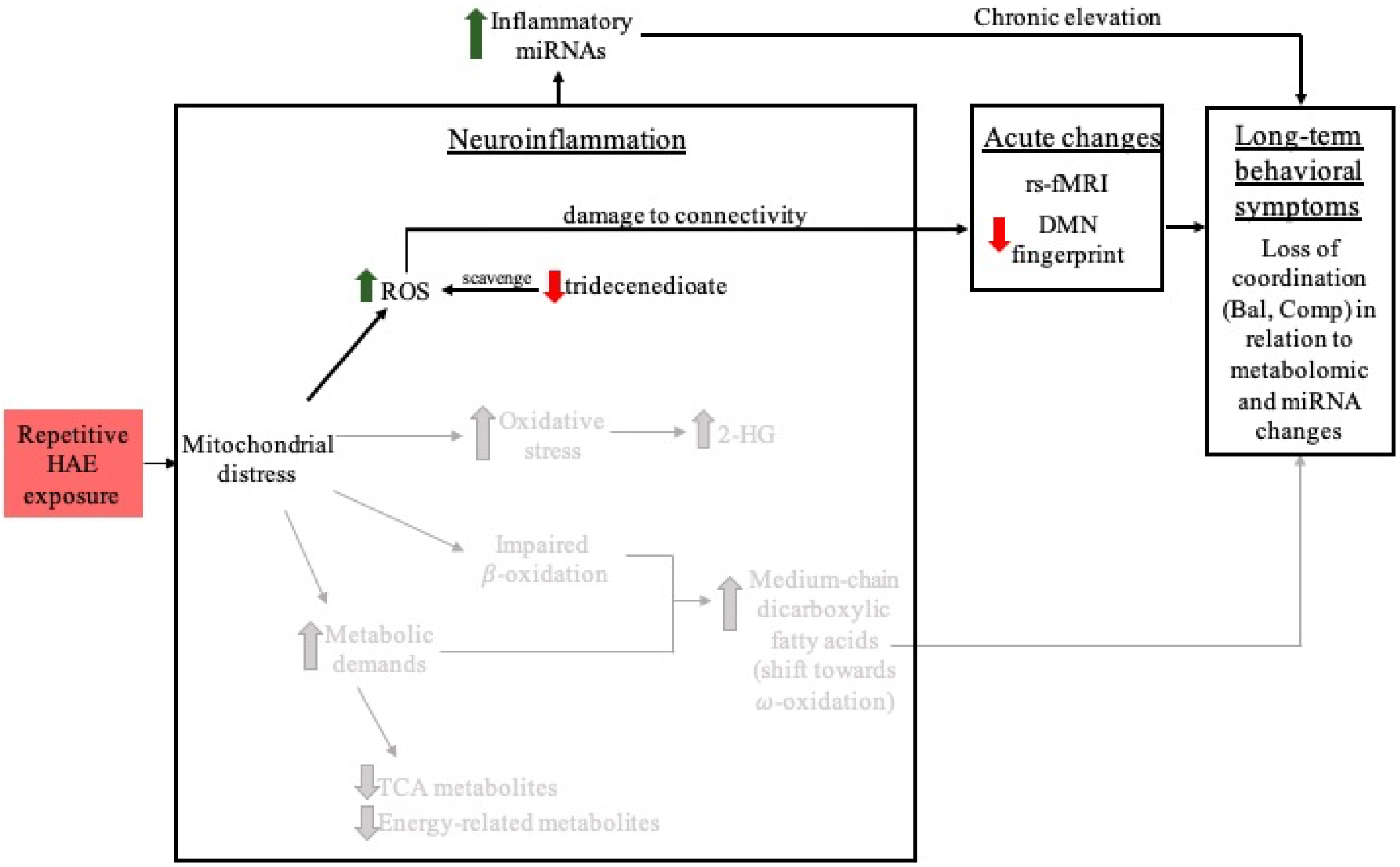

**Figure 4.** Synopsis figure summarizing the interplay between moderation variables. Repetitive head accelerate events (HAEs) in football athletes have been related to changes in brain homeostasis, as evidenced by neuroimaging studies [9,11–16,19]. Additionally, a study using the same collegiate football dataset [84] pointed to mitochondrial distress and subsequent energy imbalance, as evidenced by increased medium-chain fatty acids and decreased TCA metabolites. Furthermore, the metabolites were observed to mediate the relationship between elevated miRNAs and Luria behavior. In the present study, significant moderation effects were observed with default mode network (DMN) fingerprint, miR-505, VR balance task performance, and tridecenedioate. Tridecenedioate, a dicarboxylic fatty acid with one double bond, significantly decreased from pre- to post-season. The double bond in this metabolite is hypothesized to act as a reactive oxygen species (ROS) scavenger. Indeed, previous research supports an increase in reactive oxygen species (ROS) following repetitive exposure to head impacts [166]; therefore, a depletion of tridecenedioate would support its role as an ROS scavenger. Even with scavenging, ROS can still remain elevated and cause damage to neurophysiological systems [166]. This may ultimately result in damage to neuronal connectivity as observed by decreased similarity in DMN and interactions with VR balance task performance. Together, these complex relationships may explain why behavioral changes in subconcussed athletes are not consistently observed, but how repetitive, long-term exposure to HAEs, chronic increases of inflammatory-miRNAs, and acute changes to resting state networks could result in behavioral disturbances later in life

## Supplementary Material

### Age Matched Non-Athletes (NAth)

A cohort of sixteen undergraduate and graduate students (age: mean $\pm$ st. deviation = 22.8$\pm$ 2.12 years; nine female and seven male) from Purdue University participated in imaging session and served as Age Matched Non-Athletes (NAth) for this study.

The imaging session conducted using a 3T General Electric Signa HDx and a 16- channel brain array (Nova Medical) consisted of high resolution $T_1$ scan and a ten minutes eyes-open resting-state fMRI scan. The imaging parameters for rs-fMRI consisted of blipped echo-planar imaging: TR/TE = 2000/26 msec; flip angle = 35°; 34 slices; acceleration factor = 2; Field of View = 20 cm; voxel size = 3.125 x 3.125 x 3.80 mm and 294 volumes. The details of this imaging data are described in Bari et al., 2019 [81]. This dataset was processed using the steps listed in the Methods section and as described in [81]. Network fingerprint comparisons were conducted between age-matched non-athletes and football athletes using two-sample t-test following Bonferroni correction,

### Permutation-based Mediation Analysis

Mediation seeks to clarify the causal relationship between the independent variable (IV) and dependent variable (DV) with the inclusion of a third variable mediator (M). Mediation model proposes that instead of a direct causal relationship between IV and DV, the IV influences M which then influences the DV. Beta coefficients (β) and standard error (se) terms from the following linear regression equations are used to calculate the Sobel *p*-value and mediation effect percentage ($T_{eff}$) using the following steps:

$$Step\ 1\ (Path\ A){:}\ M = \beta_0 + \ \beta_{1A}(IV) + \epsilon_A$$

$$Step\ 2\ (Path\ B){:}\ DV = \beta_0 + \ \beta_{1B}(M) + \epsilon_B$$

$$Step\ 3\ (Path\ C,\ model\ 1){:}\ \ DV = \ \beta_0 + \ \beta_{1,1C}(IV) + \epsilon_{1C}$$

$$Step\ 4\ (Path\ C,\ model\ 2){:}\ \ DV = \ \beta_0 + \ \beta_{1,2C}(IV) + \ \beta_{2,2C}(M) + \epsilon_{2C}$$

Sobel's test is used to test if $\beta_{1,2C}$ was significantly lower than $\beta_{1,1C}$ using the following equation:

$$(3)\ Sobel\ z - score = \ \frac{(\beta_{1,1C} - \beta_{1,2C})}{\sqrt{\left[(\beta_{2,2C})^2(\epsilon_{1A})^2\right] + [(\beta_{1A})^2(\epsilon_{2C})^2]}}$$

Using a standard 2-tail z-score table, the Sobel *p*-value is determined from Sobel z-score. Mediation effect percentage $T_{eff}$ is calculated using the following equation:

$$(4)\ T_{eff} = 100 * \frac{(\beta_{1A} * \beta_{2,2C})}{(\beta_{1A} * \beta_{2,2C}) + [\beta_{1,1C} - (\beta_{1A} * \beta_{2,2C})]}$$

For this study, permutation-based mediation analysis was performed for all three-way associations following the steps listed below:

1. Mediation analysis was performed by assigning the original data variables $\Delta\boldsymbol{x_i}, \Delta\boldsymbol{y_j}, \Delta\boldsymbol{z_k}$ as IV, DV and M to obtain reference Sobel z-score: $z_0$ and $T_{eff}$. Only variables that formed three-way associations were considered.
2. Data permutation: values were randomly selected from $\boldsymbol{x}_{1,i}$ and $\boldsymbol{x}_{2,i}$ to assign to $\boldsymbol{x'}_{1,i}$ and $\boldsymbol{x'}_{2,i}$.
3. Across season measures were computed from the permuted dataset $\Delta\boldsymbol{x}'_i = \boldsymbol{x}'_{2,i} - \boldsymbol{x}'_{1,i}$. Similarly, $\Delta\boldsymbol{y}'_j$ and $\Delta\boldsymbol{z}'_k$ were computed.
4. Mediation analysis was performed on the permuted dataset $\Delta\boldsymbol{x}'_i$, $\Delta\boldsymbol{y}'_j$, $\Delta\boldsymbol{z}'_k$ by assigning as IV, DV and M and the test statistics: $z'_q$ was obtained.
5. The counter variable $K$ was incremented by one if absolute value of $z_0$ was greater than absolute value of $z'_q$.
6. Steps 2-5 were repeated: $q = 1,2, \cdots, Q$ times.
7. Permutation-based $p$-value $p_{Sobel}^{perm}$ was calculated as the proportion of the $z'_q$ values that are as extreme or more extreme than $z_0$ i.e. $K/Q$ .
8. Mediation analysis was considered significant if $p_{Sobel}^{perm} \leq 0.05$ and $T_{eff} > 50\%$.

## Results

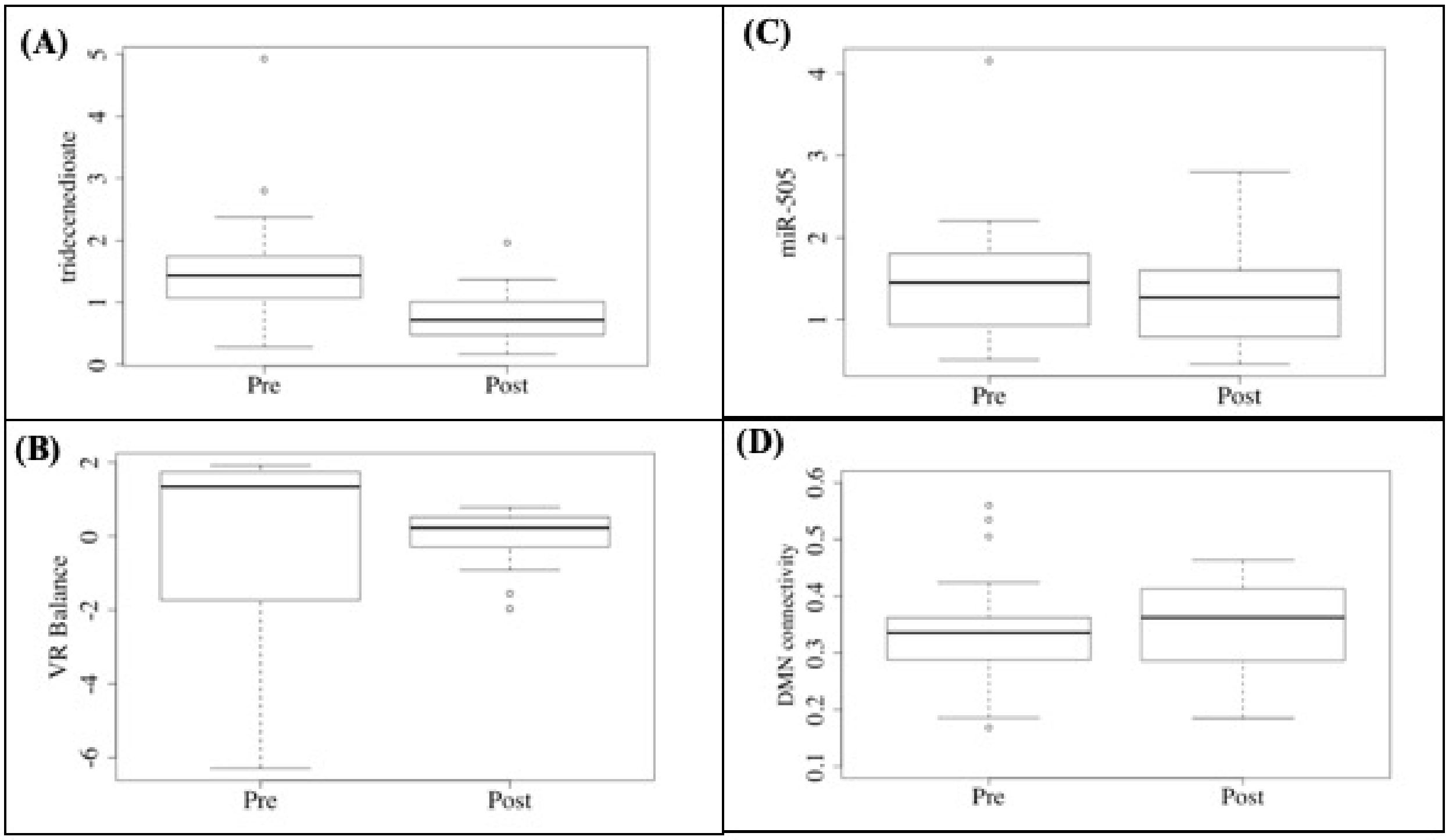

**Supplemental Figure 1**. Boxplots comparing the Pre- and Post-season distributions of (A) tridecenedioate (B) VR Balance (C) miR-505 and (D) DMN connectivity.

### (A)

| ΔVR task (Y) | Network(Z) | Cook's Outliers | β | *p* | $R^2_{adj}$ |
|---|---|---|---|---|---|
| Comprehensive | SM | 2/20 | -0.684 | 0.010 | 0.311 |
| Comprehensive | DA | 1/20 | -0.570 | 0.049 | 0.162 |
| Balance | FC | 1/20 | 0.439 | 0.020 | 0.222 |
| Balance | FP | 1/20 | 0.538 | 0.020 | 0.232 |
| Balance | DMN | 1/20 | 0.447 | 0.020 | 0.240 |
| Balance | SUBC | 1/20 | 0.430 | 0.030 | 0.202 |
| Reaction Time | Vis | 1/20 | -0.642 | 0.030 | 0.193 |
| Reaction Time | SM | 1/20 | -0.745 | 0.001 | 0.473 |
| Reaction Time | DA | 1/20 | -0.702 | 0.001 | 0.466 |

### (B)

| ΔMetabolite (X) | Network(Z) | Cook's Outliers | β | *p* | $R^2_{adj}$ |
|---|---|---|---|---|---|
| 2-hydroxyglutarate | SM | 2/20 | -0.427 | 0.040 | 0.181 |
| 1,7-dimethylurate | SM | 1/20 | -0.393 | 0.025 | 0.217 |
| 7-hydroxyoctanate | FC | 1/20 | 0.475 | 0.030 | 0.208 |
| tridecenedioate | Vis | 1/20 | -0.352 | 0.013 | 0.274 |
| tridecenedioate | DMN | 2/20 | -0.261 | 0.048 | 0.174 |

**(C)**

| ΔmiRNA (Y) | Network(Z) | Cook's Outliers | β | *p* | $R^2_{adj}$ |
|---|---|---|---|---|---|
| 505 | DMN | 1/17 | -0.458 | 0.020 | 0.302 |
| 92a | DMN | 1/17 | -0.365 | 0.015 | 0.306 |

**(D)**

| ΔVR task (Y) | ΔmiRNA (X) | Cook's Outliers | β | *p* | $R^2_{adj}$ |
|---|---|---|---|---|---|
| Comprehensive | 505 | 1/20 | -0.671 | 0.003 | 0.387 |
| Comprehensive | 30d | 0/20 | -0.586 | 0.007 | 0.306 |
| Comprehensive | 92a | 0/20 | -0.479 | 0.033 | 0.186 |
| Comprehensive | 151-5p | 1/20 | -0.670 | 0.002 | 0.395 |
| Balance | 505 | 1/20 | -0.533 | 0.003 | 0.389 |
| Balance | 30d | 1/20 | -0.417 | 0.027 | 0.213 |
| Reaction Time | 20a | 0/20 | -0.458 | 0.043 | 0.184 |
| Reaction Time | 505 | 1/20 | -0.569 | 0.016 | 0.254 |
| Reaction Time | 30d | 0/20 | -0.483 | 0.031 | 0.190 |
| Reaction Time | 92a | 0/20 | -0.537 | 0.015 | 0.249 |
| Reaction Time | 151-5p | 1/20 | -0.659 | 0.007 | 0.323 |

**(E)**

| ΔVR task (Y) | ΔMetabolite (X) | Cook's outliers | β | *p* | $R^2_{adj}$ |
|---|---|---|---|---|---|
| Comprehensive | 2-hydroxyglutarate | 2/23 | 0.622 | 0.008 | 0.279 |
| Comprehensive | corticosterone | 2/23 | -0.748 | 0.008 | 0.281 |
| Comprehensive | caffeine | 1/23 | 0.575 | 0.052 | 0.134 |
| Comprehensive | 1,7-dimethylurate | 1/23 | 0.568 | 0.035 | 0.165 |
| Comprehensive | sebacate | 2/23 | 0.647 | 0.002 | 0.373 |
| Comprehensive | dodecadienoate | 3/23 | 0.458 | 0.045 | 0.160 |
| Comprehensive | azelate | 1/23 | 0.475 | 0.022 | 0.196 |
| Comprehensive | suberate | 1/23 | 0.514 | 0.017 | 0.214 |
| Comprehensive | 8-hydroxyoctanate | 1/23 | 0.561 | 0.007 | 0.276 |
| Comprehensive | heptanoate (7:0) | 1/23 | 0.632 | 0.011 | 0.248 |
| Comprehensive | 1 2 3-benzenetriol sulfate (2) | 1/23 | 0.769 | 0.029 | 0.179 |
| Comprehensive | adenosine | 0/23 | -0.609 | 0.002 | 0.341 |
| Balance | azelate | 1/23 | 0.386 | 0.031 | 0.174 |
| Balance | 7-hydroxyoctanate | 2/23 | 0.398 | 0.040 | 0.162 |
| Balance | 8-hydroxyoctanate | 2/23 | 0.579 | 0.004 | 0.335 |
| Balance | undecanedioate | 1/23 | 0.356 | 0.048 | 0.140 |
| Balance | heptanoate (7:0) | 1/23 | 0.387 | 0.033 | 0.167 |
| Balance | tridecenedioate | 1/23 | -0.697 | 0.041 | 0.152 |
| Reaction Time | alpha-keroglutarate | 2/23 | 0.493 | 0.027 | 0.193 |
| Reaction Time | 2-hydroxyglutarate | 2/23 | 0.576 | 0.011 | 0.255 |
| Reaction Time | caffeine | 1/23 | 0.654 | 0.024 | 0.193 |
| Reaction Time | adenosine | 1/23 | -0.642 | 0.004 | 0.317 |

**(F)**

| ΔmiRNA (Y) | ΔMetabolite (X) | Cook's Outliers | β | *p* | |
|---|---|---|---|---|---|
| 20a | azelate | 1/20 | -0.479 | 0.012 | 0.309 |
| 20a | undecanedioate | 1/20 | -0.393 | 0.042 | 0.197 |
| 505 | sebacate | 2/20 | -0.528 | 0.010 | 0.309 |
| 505 | azelate | 2/20 | -0.612 | 0.003 | 0.404 |
| 505 | suberate | 2/20 | -0.542 | 0.008 | 0.320 |
| 505 | 8-hydroxyoctanate | 2/20 | -0.529 | 0.041 | 0.188 |
| 505 | undecanedioate | 2/20 | -0.555 | 0.008 | 0.326 |
| 505 | heptanoate (7:0) | 2/20 | -0.425 | 0.029 | 0.218 |
| 505 | tridecenedioate | 1/20 | 0.256 | 0.050 | 0.156 |
| 3623p | phosphate | 2/20 | -0.358 | 0.054 | 0.163 |
| 3623p | linoleate3n6 | 3/20 | 0.339 | 0.037 | 0.210 |
| 3623p | heptanoate (7:0) | 0/20 | -0.543 | 0.013 | 0.256 |
| 30d | sebacate | 0/20 | -0.528 | 0.017 | 0.239 |
| 30d | heptanoate (7:0) | 1/20 | -0.622 | 0.005 | 0.346 |
| 30d | O-sulfo-L-tyrosine | 0/20 | -0.436 | 0.055 | 0.145 |
| 30d | adenosine | 2/20 | 0.571 | 0.018 | 0.259 |
| 92a | sebacate | 2/20 | -0.615 | 0.001 | 0.510 |
| 92a | azelate | 4/20 | -0.642 | 0.003 | 0.437 |
| 92a | suberate | 3/20 | -0.716 | 0.008 | 0.346 |
| 92a | undecanedioate | 2/20 | -0.611 | 0.004 | 0.375 |
| 92a | heptanoate (7:0) | 0/20 | -0.523 | 0.018 | 0.234 |
| 92a | O-sulfo-L-tyrosine | 0/20 | -0.506 | 0.023 | 0.215 |
| 92a | adenosine | 2/20 | 0.457 | 0.036 | 0.199 |
| 195 | sebacate | 1/20 | -0.493 | 0.016 | 0.255 |
| 195 | suberate | 3/20 | -0.399 | 0.053 | 0.175 |
| 195 | 8-hydroxyoctanate | 1/20 | -0.567 | 0.004 | 0.364 |
| 195 | undecanedioate | 2/20 | -0.474 | 0.020 | 0.238 |
| 195 | heptanoate (7:0) | 1/20 | -0.608 | 0.002 | 0.412 |
| 195 | O-sulfo-L-tyrosine | 1/20 | -0.740 | 0.020 | 0.238 |
| 93p | caffeine | 2/20 | -0.554 | 0.024 | 0.235 |
| 93p | 7-hydroxyoctanate | 1/20 | 0.757 | 0.018 | 0.018 |
| 1515p | stearidonate | 3/20 | 0.499 | 0.020 | 0.263 |
| 1515p | 8-hydroxyoctanate | 1/20 | -0.423 | 0.038 | 0.185 |
| 1515p | heptanoate (7:0) | 2/20 | -0.560 | 0.015 | 0.273 |

**(G)**

| Network (Y) | HAE (X) | Cook's outliers | β | *p* | $R^2_{adj}$ |
|---|---|---|---|---|---|
| Vis | $aHAE_{80G}$ | 1/20 | -0.150 | 0.019 | 0.242 |
| SM | sess | 0/20 | 0.006 | 0.032 | 0.187 |
| FP | $aHAE_{25G}$ | 2/20 | -0.031 | 0.026 | 0.227 |
| DMN | $cHAE_{80G}$ | 1/20 | -0.009 | 0.031 | 0.201 |
| SUBC | $cHAE_{80G}$ | 1/20 | -0.016 | 0.001 | 0.452 |
| SUBC | $aHAE_{25G}$ | 1/20 | -0.030 | 0.009 | 0.302 |
| SUBC | $aHAE_{80G}$ | 1/20 | -0.438 | 0.001 | 0.432 |

(H)

| ΔVR task (Y) | HAE (X) | Cook's outliers | β | *p* | $R^2_{adj}$ |
|---|---|---|---|---|---|
| Reaction Time | $cHAE_{25G}$ | 1/23 | -0.582 | 0.007 | 0.278 |

**(I)**

| ΔmiRNA (Y) | HAE (X) | Cook's outliers | β | *p* | $R^2_{adj}$ |
|---|---|---|---|---|---|
| 505 | $cHAE_{25G}$ | 1/20 | 0.397 | 0.035 | 0.191 |
| 362-3p | $cHAE_{80G}$ | 3/20 | -0.488 | 0.045 | 0.191 |
| 92a | $aHAE_{80G}$ | 2/20 | 0.882 | 0.046 | 0.178 |

**(J)**

| ΔMetabolite (Y) | HAE (X) | Cook's outliers | β | *p* | $R^2_{adj}$ |
|---|---|---|---|---|---|
| fumrate | $aHAE_{80G}$ | 1/23 | -0.687 | 0.031 | 0.173 |
| linolenate3n6 | $cHAE_{80G}$ | 3/23 | 0.327 | 0.038 | 0.175 |
| sebacate | $cHAE_{25G}$ | 1/23 | 0.571 | 0.007 | 0.279 |
| sebacate | $aHAE_{25G}$ | 2/23 | 0.454 | 0.028 | 0.190 |
| dodecadienoate | sess | 1/23 | 0.497 | 0.014 | 0.229 |
| suberate | $cHAE_{25G}$ | 2/23 | 0.412 | 0.034 | 0.174 |
| suberate | $aHAE_{25G}$ | 2/23 | 0.440 | 0.032 | 0.178 |
| undecanedioate | $cHAE_{25G}$ | 1/23 | 0.561 | 0.008 | 0.264 |
| undecanedioate | $aHAE_{80G}$ | 2/23 | 0.461 | 0.030 | 0.183 |
| carboxyethylvaline | $aHAE_{25G}$ | 1/23 | 0.419 | 0.025 | 0.190 |
| mannose | sess | 1/23 | 0.496 | 0.016 | 0.221 |
| O-sulfo-L-tyrosine | sess | 1/23 | 0.514 | 0.005 | 0.300 |

**Supplemental Table 1**. Tables (A-J) lists all pairwise nominal interactions at a significance level of 0.05 between the variables: Network fingerprint, ΔMetabolite, ΔmiRNA, ΔVR tasks and HAE metrics. Cook's outliers (*k*/*n*) lists the number of outliers *k*, based on Cook's distance, removed out of the total *n* samples. The linear regression slopes β, *p*-values and Adjusted $R_2$ ($R_{2adj}$) are reported after the outlier removal.